# A Modeling Approach of Return and Volatility of Structured Investment Products with Caps and Floors

by

Jiaer He[1], Roberto Rivera[2]

[1] College of Business, University of Puerto Rico-Mayaguez
[2] Department of Mathematical Sciences, University of Puerto Rico-Mayaguez

## Abstract

Popular investment structured products in Puerto Rico are stock market tied Individual Retirement Accounts (IRA), which offer some stock market growth while protecting the principal. The performance of these retirement strategies has not been studied. This work examines the expected return and risk of Puerto Rico stock market IRA (PRIRA$_s$) and compares their statistical properties with other investment instruments before and after tax. We propose a parametric modeling approach for structured products and apply it to PRIRAs. Our method first estimates the conditional expected return (and variance) of PRIRA assets from which we extract marginal moments through the Law of Iterated Expectation. Our results indicate that PRIRA$_s$ underperform against investing directly in the stock market while still carrying substantial risk. The expected return of the stock market IRA from Popular Bank (PRIRA$_1$) after tax is slightly greater than that of investing in U.S. bonds, while PRIRA$_1$ has almost two times the risk. The stock market IRA from Universal (PRIRA$_2$) performs similarly to PRIRA$_1$, while PRIRA$_2$ has a lower risk than PRIRA$_1$. PRIRA$_s$ may be reasonable for some risk-averse investors due to their principal protection and tax deferral.

## Introduction

A structured investment product, also named a market-linked investment, is one of the most popular financial derivatives. The payoff at maturity relies on one or more specified assets (mostly stocks or stock indices). Because structured products emphasize downside protection with simultaneous participation in the upside, they are very attractive to retail investors (Blundell-Wignall, 2007). However, banks may design complex products to cater to yield-seeking households (Bordalo, Gennaioli, & Shleifer, 2016). Based on the principal-protected note (PPN) principle, some Puerto Rican institutions create pre-packaged individual retirement accounts with returns tied to the stock market.

Retirement products such as PRIRA are not common in places other than Puerto Rico, and indeed, no evidence of them elsewhere was found. Moreover, little is known about the statistical properties of these assets. Puerto Rico has been going through an economic recession since 2006. Leaders on the island have been attempting to find ways to boost the economy (Rivera, 2016) and data-driven methods may help improve investment opportunities (Rivera and Rolke, 2019; Rivera, R., Marazzi, M., & Torres-Saavedra, P. A. 2019; Rivera, R., Rosenbaum, J. E., & Quispe, W., 2020; Rivera, 2020; Rivera, R., & Rosenbaum, J. 2020; Rosenbaum, J. E., Stillo, M., Graves, N., & Rivera, R. 2021; Lugo, O., & Rivera, R. 2023). Since 2013, the government has been forced to reform its deficient public-employee pensions. Under the pension reforms, there was a noticeable increase in the retirement age and employee contributions (Austin, 2016). However, most Puerto Ricans seem to be unprepared to take charge of financial decisions and retirement (Castro-Gonzales, 2014). Therefore, it is important to get people to realize the significance of building good retirement savings plans.

Although research on different aspects of structured products exists, no prior studies were found that have investigated the structured products tied to Puerto Rican IRA and the statistical properties they present. As a structured product, PRIRA$_s$ not only provide principal protection but also come with the usual tax benefits of individual retirement accounts. The Puerto Rico Investment Companies Act, which was approved in 1994, states that 67% of the capital of an investment company should be invested in local securities. Although this law is intended to promote economic growth, the limited options for local companies to invest greatly hinder the local investment companies' options to construct their investment instruments. This makes it difficult for investment companies to construct investment alternatives. Local investment products such as PRIRA$_s$ may have some specific properties which are not common outside Puerto Rico. The importance of this study is to examine the expected return and risk traits of the PRIRA$_s$ before- and after-tax and make some comparisons with other investment instruments. The findings of this study will help guide a lot of future decisions made by investors and pundits.

Puerto Rico IRA systems, both traditional and Roth plans, are very different from U.S. IRAs. Income sourced from non-federal employment within Puerto Rico cannot be contributed to a U.S. IRA. Notably, the U.S. IRA usually is presented in the form of an empty box. Most investors can place anything (stocks, bonds, index funds, etc.) up to $6,000 (currently) in the IRA. As a result, they are exposed to investment risks. In contrast, in Puerto Rico, IRAs are a prepackaged investment instrument. It either comes with fixed returns, for example, using CDs, or may offer returns contingent on something like the S&P 500. In this study, we examine two IRA products, one from Popular Bank, and the other from Universal insurance company. Both are the typical structured products whose returns are determined by the performance of the S&P 500 index. The invested principal is protected by issuers.

The PRIRAs are Structured Certificates of Deposit (SCDs), whose payoffs depend on the performance of the S&P 500 index. If the underlying asset does well, the payoff from the SCD will be limited to a predetermined cap, which is specified by a particular SCD. However, the SCDs can guarantee to return the principal at maturity date even though the underlying financial assets perform poorly. In particular, the SCDs are typically worth about 93% of the value of a contemporaneously issued fix-rate CD (Deng, Dulaney, Husson, & McCann, 2013). Generally, the SCDs are constructed by a risk-free asset, which is combined with Asian call options to obtain the desired payoff. $PRIRA_1$ is made up of the following components: (i) A zero-coupon CD issued at a discount to par value with a five-year maturity. Popular Bank can get funding from issuing this product. The cost of funds will depend on the discount and its implied accretion to par; (ii) a long call Asian option at the money on the S&P 500 (strike price = 100% of the initial index value). The option will be a cost to the bank. (iii) a short call Asian option out of the money on the S&P 500 (strike price = 125% of the initial index value). The option will provide a credit to the bank. This option effectively caps the upside the client will receive because he/she purchases the 100% call option and sells the 125% call option. The information about how Universal constructs the $PRIRA_2$ is considered privileged and is not shared with the public yet we believe it works similarly to $PRIRA_1$.

## Volatility Model

We model the expected return and risk of each of the $PRIRA_i$, applying the truncated normal distribution and the iterated expectation model. The latter incorporates the variance of the returns in the estimation. We will compare the statistical properties of $PRIRA_i$ with other investment instruments, such as deposit savings, an index fund that follows the S&P 500, a bond index fund, and a diversified portfolio comprising of stocks and bonds. We account for tax effects on the expected return and risk.

Modeling Return and Volatility of PRIRAs

We can apply the law of Iterated Expectation (Weiss, Holmes, & Hardy, 2005) to determine the expected return of PRIRA₁ and PRIRA₂. The return of the S&P 500 is a continuous random variable, while the return of each PRIRA, $Y_i$, follows a mixture distribution (Lovric, 2011).

To calculate the expected value for the PRIRA₁ for the three different segments, we can multiply each possible outcome, y, by its corresponding probability, $P(Y_1=y|G_1)$. However, $P(Y_1=0|G_1 \leq 0)$ is equal to one. It follows in a similar fashion with $P(Y_1=0.5|G_1 \geq 0.4)$. We can calculate each expected return for the three different segments. Thus, we have:

Let: $G_1$ = Average return in S&P 500 over 5-years

$$G_1 = \frac{Average\ Index\ Value - Initial\ Index\ Value}{Initial\ Index\ Value}$$

$$E[Y_1|G_1] = \begin{cases} 0, & G_1 \leq 0 \\ 1.25 * E[G_1|0 < G_1 < 0.4], & 0 < G_1 < 0.4 \\ 0.50, & G_1 \geq 0.4 \end{cases} \quad (1)$$

where:

$$E[Y_1] = E[E[Y_1|G_1]] = 0 * P(G_1 \leq 0) + 1.25 * E[G_1|0 < G_1 < 0.4] * P(0 < G_1 < 0.4) + 0.5 * P(G_1 \geq 0.4) \quad (2)$$

Let: $G_2$ = Index Return in S&P 500 for any year

$$G_2 = \frac{Current\ Anniversary\ Index\ Value - Prior\ Anniversary\ Index\ Value}{Prior\ Anniversary\ Index\ Value}$$

Similar to PRIRA₁, we can calculate each expected return of PRIRA₂ as:

$$E[Y_2|G_2] = \begin{cases} 0, & G_2 \leq 0 \\ E[G_2|0 < G_2 < 0.2025], & 0 < G_2 < 0.2025 \\ 0.2025, & G_2 \geq 0.2025 \end{cases}$$

where:

$$E[Y_2] = E[E[Y_2|G_2]] = 0 * P(G_2 \leq 0) + E[G_2|0 < G_2 < 0.2025] * P(0 < G_2 < 0.2025) + 0.2025 * P(G_2 \geq 0.2025) \quad (3)$$

Marginal and conditional probabilities for $G_i$, i=1, 2, will be modeled with the normal and truncated normal distributions, respectively (Burkardt, 2014).

Similarly, in calculating the variance of the return of each PRIRA, the law of Total Variance (Weiss, Holmes, & Hardy, 2005) is applied to calculate the variance of PRIRA₁ as:

$$Var(Y_1) = E[Var(Y_1|G_1)] + Var(E[Y_1|G_1])$$

We have:

$$Var(Y_1|G_1) = \begin{cases} 0, G_1 \leq 0 \\ 1.25^2 * Var(G_1|0 < G_1 < 0.4), 0 < G_1 < 0.4 \\ 0, G_1 \geq 0.4 \end{cases} \quad (4)$$

Then

$$E[Var(Y_1|G_1)]$$
$$= 0 * P(G_1 \leq 0) + 1.25^2 * Var(G_1|0 < G_1 < 0.4)$$
$$* P(0 < G_1 < 0.4) + 0 * P(G_1 \geq 0.4) \quad (5)$$

And

$$Var(E[Y_1|G_1])$$
$$= (0 - E[Y_1])^2 * P(G_1 \leq 0)$$
$$+ (E[Y_1|0 < G_1 < 0.4] - E[Y_1])^2 * P(0 < G_1 < 0.4)$$
$$+ (0.5 - E[Y_1])^2 * P(G_1 \geq 0.4) \quad (6)$$

Because $P(Y_1=0|G_1\leq 0)=1$, $Var(Y_1=0|G_1\leq 0)=0$. Also, $P(Y_1=0.5|G_1\leq 0.4)=1$, $Var(Y_1=0.5|G_1\leq 0.4)=0$. $Var(Y_1|0<G_1<0.4)$ can be determined by the fact that $G_1$ given $0<G_1<0.4$ follows a truncated normal distribution.

The variance of $PRIRA_2$ is obtained similar to the variance of $PRIRA_1$.

In this study, we apply the normal and truncated normal distribution to build the return and volatility models for PRIRAs. The Shapiro-Wilk test is used to examine the normal distribution of the S&P 500 returns.

### Comparison Benchmarks

We aim to compare the performances of $PRIRA_s$ with other investment approaches to determine which investment is the best one for the investor's needs. Specifically, we choose another four common investment instruments to make comparisons: deposit savings; an index fund that follows the S&P 500; a bond market index fund; a diversified portfolio of stocks and bonds.

We will use proxies of these investment strategies. Current deposit account interest rates in Puerto Rico will be obtained to model returns from deposit savings, Vanguard 500 Index Fund Investor Shares (VFINX) will represent index fund, Vanguard Total Bond Market Index Fund Investor Shares (VBMFX) will represent bond market fund, and the diversified portfolio (of Vanguard stock and bond index funds) will represent a diversified portfolio. We assume the annual expected return of a savings account is 0.25%. For the annual expected return and risk for index fund, bond market fund, and diversified portfolio, we will use the sample mean and sample standard deviation, respectively.

We assume investment horizons of 10 and 30-years. To ease the comparison of the return for each strategy, we arbitrarily assume that the investor invests $10,000 per year. Note that for

the IRA, this implies that the investment strategy is for a couple who files jointly. We use the 10-yr U.S. T-bond yield as the risk-free rate to calculate the Sharpe ratio. The average 10-yr U.S. T-bond over all periods is about 5.96%[1]

## Tax Adjustment

We need to take the tax effects into consideration to measure the portfolio value of each investment strategy. Puerto Rico, as a territory of the United States, has its own tax system, which is accommodated by provisions of US law. The government imposes a local Puerto Rico income tax instead of the typical U.S. federal income tax. Puerto Rican residents are subject to taxes in Puerto Rico on their worldwide income, no matter where the income is sourced. Table 1 lists the tax rates for an individual's ordinary income for 2022 in Puerto Rico.[2]

**Table 1 Tax Bracket for Ordinary Income for 2022 in Puerto Rico**

| Net taxable income | Tax rate |
|---|---|
| Up to $9,000 | 0% |
| More than $9,000 but not more than $25,000 | 7% |
| More than $25,000 but not more than $41,500 | 14% |
| More than $41,500 but not more than $61,500 | 25% |
| More than $61,500 | 33% |

Under the provisions of the Internal Revenue Code for Puerto Rico, approved on January 31, 2011, an individual receiving dividends from a local Puerto Rican company must pay a 15% special tax. If the dividends received are from a foreign corporation, these dividends are taxed at the ordinary income tax rate. Interest income from deposits in local banking institutions are subject to a specific 10% tax. When we consider the tax effects on the investment instruments in this work, we focus on married couples who elect to file jointly.

---

[1] Source: http://people.stern.nyu.edu/adamodar/New_Home_Page/datafile/histretSPX.html

[2] Source: The Internal Revenue Code of Puerto Rico

For simplicity, we assume no withdrawals during the investment period. Furthermore, we will model the traditional IRA through the PRIRAs, whose contributions are tax-deductible only if the money is placed in the account. In other words, the portfolio value of PRIRAs will be the same as that of PRIRAs before tax.

For VFINX and VBMFX, investors should pay the dividend tax every year and defer paying the capital gains tax until they sell assets. In this work, we assume that the investor exploits the compound interest effect to the fullest by not withdrawing money during the whole investing period. Thus, we just need to consider the dividend tax. After the investment period, the investor must pay capital gain taxes, as investors who own PRIRAs must. Dividends, belonging to the exotic dividends, are taxed at the ordinary income tax rates. In this study, we will examine the expected return and expected portfolio value after tax depending on the different tax brackets. For the diversified portfolio, we assume the investor invests in VFINX as the stocks and VBMFX as the bonds separately based on the different weights. The dividends from the VFINX and VBMFX are taxed according to normal income tax rates. Similar to index fund and bond market index fund, there is no need to consider the capital gain tax in our analysis.

### Data

This study includes the S&P 500 index value. Specifically, we used the closing prices of the S&P 500 as a measure of the stock market value. Moreover, we recorded the annual return rate and dividend return rate of the stock market index fund and the bond market index fund. All the data was downloaded from Yahoo! Finance.

Closing prices of the S&P 500 were used to establish the initial index value and monthly end index value from April 1957 to March 2019 to calculate $PRIRA_1$ returns. When analyzing $PRIRA_1$, we arbitrarily assumed that the opening date was April 2$^{nd}$ of each year. The closing prices of the S&P 500 on every April 1$^{st}$ from 1957 to 2014 were collected and noted as each initial index value. We also collected the closing price of the S&P 500 on the last business day of each month during the term as the monthly end index value. For $PRIRA_2$, we collected the value of the S&P 500 index of every anniversary. We assumed that the policy's first anniversary was April 1$^{st}$, 1957. The time range for $PRIRA_2$ was from 1957 to 2019.

The data employed in VFINX and VBMFX consisted of the annual total return rate to calculate the expected annual return and the risk of return from 1987 to 2019. When we consider the tax effect on VFINX and VBMFX, we just need to focus on the annual dividends return. The VFINX and the VBMFX were used as proxies to measure the performance of a diversified portfolio. We used the annual total return rate and annual dividends return of VFINX and VBMFX from 1987 to 2019, respectively.

## Results

The mean of $G_1$ from 1957 to 2019 was 0.212, which we treated as the mean 5-year return of the S&P 500 ($\mu_{G_1} = 0.212$). The standard deviation of $G_1$ was 0.24 ($\sigma_{G_1} = 0.24$). Thus:

$G_1 \sim N(0.212, 0.24)$, which was used to find required probabilities for (2), (5) and (6).

A similar approach was implemented based on $G_2 \sim N(0.258, 0.314)$

## Volatility Model Results

We applied a truncated normal distribution to PRIRA$_s$ return. Specifically, the distribution of $G_1$ given the value range from 0 to 0.4 is a truncated normal $\Psi_{G_1}(G_1; 0.212, 0.24, 0, 0.4)$. The expected value of $G_1$, given $G_1$ ranges between 0 and 0.4, is 0.203.

The variance $\gamma_{G_1}^2$ of the truncated normal distribution can also be regarded as a perturbation of the variance $\sigma_{G_1}^2$ of the normally distributed $G_1$. The variance of $G_1$, given $G_1$ ranges between 0 and 0.4, is 0.012.

Based on (1) and the expected value given $G_1$ ranges between 0 and 0.4, we have the following:

$$E(Y_1|G_1) = \begin{cases} 0, & G_1 \leq 0 \\ 0.254, & 0 < G_1 < 0.4 \\ 0.5, & G_1 \geq 0.4 \end{cases} \quad (7)$$

Based on (4) and the variance given $G_1$ ranges between 0 and 0.4, we have:

$$Var(Y_1|G_1) = \begin{cases} 0, & G_1 \leq 0 \\ 0.019, & 0 < G_1 < 0.4 \\ 0, & G_1 \geq 0.4 \end{cases} \quad (8)$$

Based on (2) and (7), the expected return of PRIRA$_1$ every 5 years is 0.259.

To do comparisons between all investment products, it is better to transform the 5-year expected return to an effective annual rate (EAR) (Bodie et al., 2014). We can relate EAR to the total return, $E(Y_1)$, over a 5-year holding period by using the following equation:

$$1 + EAR = [1 + E(Y_1)]^{1/5}$$

$$EAR = 0.047$$

Thus, the annual expected return of PRIRA$_1$ was 4.7%. Later we will compare this expected return with that of other investment alternatives.

Based on the (5), (6) and (8), Var(Y$_1$) =0.036.

Similarly, we will do the following transformation process to get the annual standard deviation for PRIRA$_1$. We can relate annual standard deviation ($\omega$) to the 5-year holding period standard deviation ($\sigma_{Y_1}$) by using the following equation:

$$\sigma_{Y_1} = \omega * \sqrt{5}$$

$$\omega = 0.085$$

Similar to G$_1$, we can summarize the mean and variance of a truncated normal for G$_2$, given G$_2$ ranges between 0 and 0.2025 are 0.106 and 0.0033 respectively.

Thus, the expected return and variance are:

$$E(Y_2|G_2) = \begin{cases} 0, G_2 \leq 0 \\ 0.106, 0 < G_2 < 0.2025 \\ 0.2025, G_2 \geq 0.2025 \end{cases} \quad (9)$$

$$Var(Y_2|G_2) = \begin{cases} 0, G_2 \leq 0 \\ 0.0033, 0 < G_2 < 0.2025 \\ 0, G_2 \geq 0.2025 \end{cases}$$

We apply the law of iterated expectations to analyze the expected value of $Y_2$. Based on (3) and (9), the expected return of PRIRA$_2$ every 3 years is 0.139. Thus, the annual expected return of PRIRA$_2$ was 4.45%. In section 4.2, we will compare this expected return with that of other investment alternatives.

Moreover, the annual standard deviation of PRIRA$_2$ was 4.9%. Note that the annual expected return of PRIRA$_1$ (4.7%) is greater than that of PRIRA$_2$ (4.45%). However, PRIRA$_1$ has a higher standard deviation compared to that of PRIRA$_2$, indicating that PRIRA$_1$ is riskier than PRIRA$_2$.

### Comparisons before Tax

Table 3 summarizes the performance of PRIRA$_s$ as well as the performance of the savings account, VFINX, VBMFX, and the diversified portfolio. We also estimate the Sharpe ratio for each investment. The highest annual return before tax was VFINX (11.9%), followed by the diversified portfolio and VBMFX with 10% and 6%, respectively. VFINX and VBMFX returns are slightly higher than the historical returns of stock and bond funds (Li, Tower, & Zhang, 2019). Thus, the diversified portfolio also has a slightly higher return than historical records indicate. The data also showed the expected returns of PRIRA$_1$ (4.7%), PRIRA$_2$ (4.45%), and, lastly, the savings account (0.25%). The expected return of VFINX is about 2.5 times higher than that of PRIRA$_s$. By comparison, the expected return of the diversified portfolio is about two times as much as that of PRIRA$_s$. Moreover, the expected return of PRIRA$_s$ is about 20% lower than that of VBMFX.

**Table 3 Comparisons of Annual Return, Standard Deviation, and Sharpe Ratio for Investments**

| Investment Alternatives | Annual Return | Standard Deviation | Sharpe Ratio |
|:---:|:---:|:---:|:---:|
| **PRIRA$_1$** | 4.7% | 8.5% | -14.8% |
| **PRIRA$_2$** | 4.45% | 4.9% | -30.82% |
| **Savings Account** | 0.25% | / | / |
| **VFINX** | 11.9% | 17% | 34.9% |
| **VBMFX** | 6% | 4.7% | 0.85% |
| **Diversified Portfolio** | 10% | 12.2% | 33.11% |

PRIRA$_s$ and VBMFX had an expected return that beat inflation with moderate standard deviations. They can be considered a viable option for risk-averse investors. However, the tax effect on VBMFX may affect the investors' decisions. We will discuss the tax effect in more detail in the next section.

The standard deviation of VFINX experienced the highest value at 17% during the whole period (Table 3). This means that the dispersion among the returns is higher than the other investing alternatives. In contrast, PRIRA$_1$ had a standard deviation of 8.5%, while for PRIRA$_2$ it was 4.9%. The volatility of VFINX is two times higher than that of PRIRA$_1$ and about 3.5 times higher than that of PRIRA$_2$. However, the volatility of PRIRA$_1$ is about 1.8 times higher than that of VBMFX, even though PRIRA$_1$ has a lower expected return than VBMFX. The standard deviation of PRIRA$_1$ is about 1.7 times higher than that of PRIRA$_2$.

The VFINX offered an average risk premium of 5.94% and a standard deviation of 17%, resulting in a reward-to-volatility ratio of 34.9%. In other words, VFINX investors enjoyed a 34.9% average excess return over the T-bond rate for every 1% of standard deviation. The Sharpe ratios of VBMFX and the diversified portfolio for the overall period were 0.85% and 33.11%, especially. Moreover, the Sharpe ratios of VFINX and the diversified portfolio indicate that the assets are in line with the broader market.

$PRIRA_1$ had a negative excess return (-1.26%) with a large standard deviation (8.5%), which made the Sharpe ratio less negative compared to $PRIRA_2$. $PRIRA_2$ had a small negative excess return associated with small volatility.

**Figure 1 The Expected Portfolio Value for Each Investment before Tax over 10 Years**

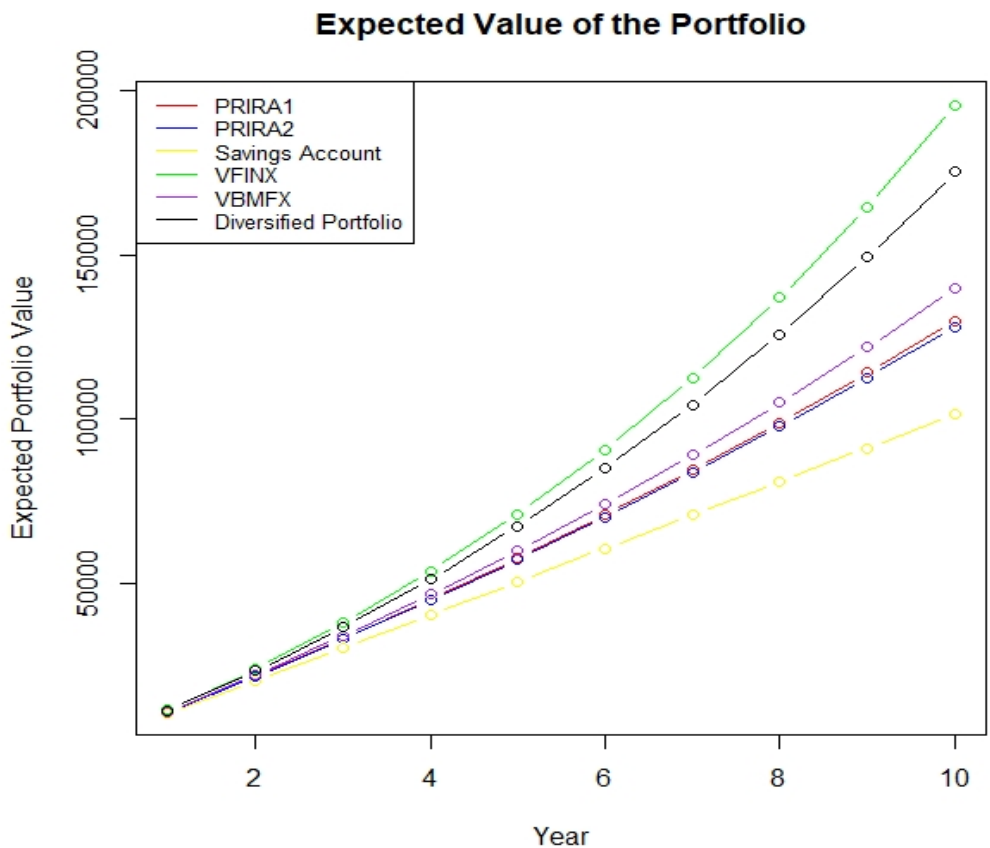

Investing on the stock market and the diversified portfolio showed the best appreciation over 10 years (Figure 4). The portfolio value of VBMFX performed better than PRIRAs and the savings account; which increased to just slightly above one hundred thousand dollars after 10 years, which is the principal invested in such a strategy.

**Figure 2 The Expected Portfolio Value for Each Investment before Tax over 30 Years**

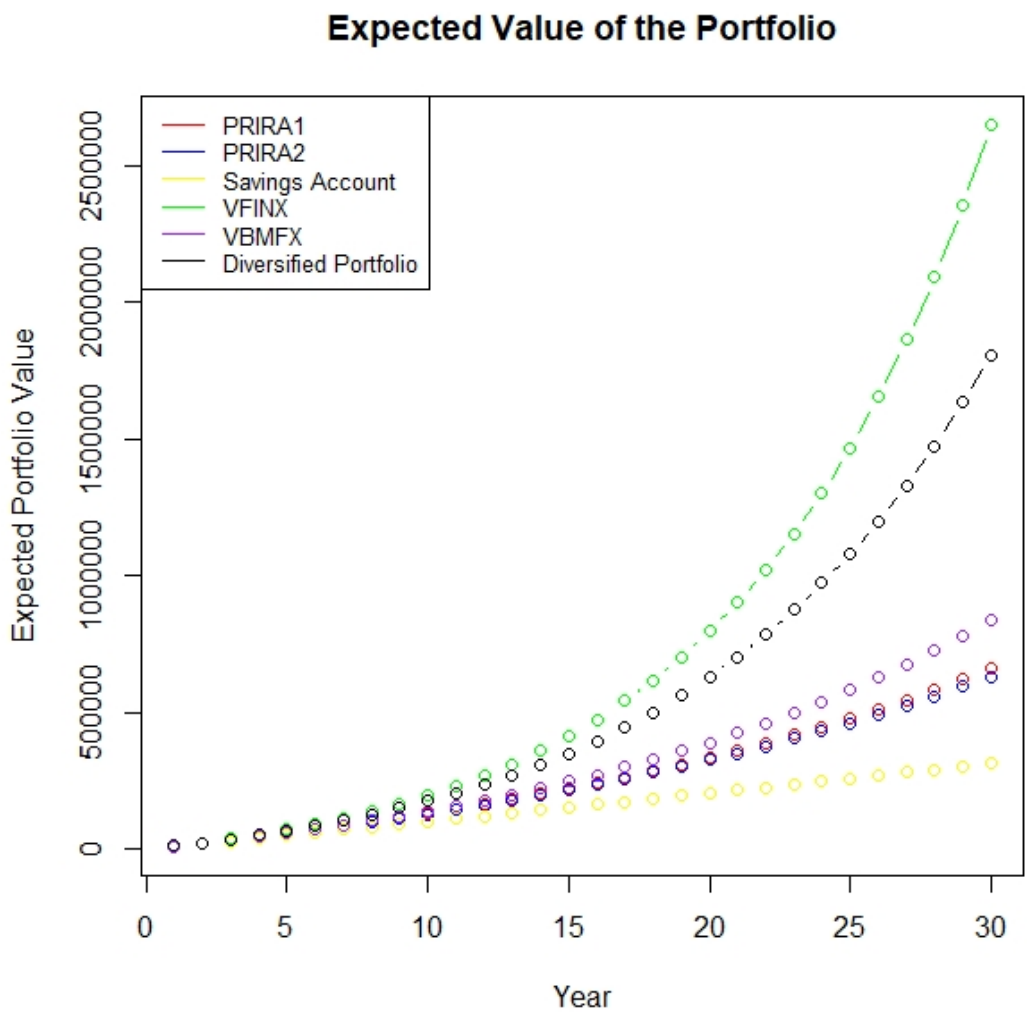

Figure 5 illustrates best the impact compound interest has on the value of each portfolio, particularly the VFINX and the diversified portfolio. There was a dramatic growth in the expected portfolio value of VFINX (green line) after 15 years and then raised rapidly to a peak of more than 2 million, which far exceed that of other investments. The expected portfolio value of the diversified portfolio continued to increase but more steeply to more than 1.5 million at the end of 30 years, relatively about 600 thousand smaller than that of VFINX. The expected portfolio values of VBMFX, PRIRAs, and the savings account remained stable during the whole period. The expected portfolio of $PRIRA_1$ did better than that of $PRIRA_2$. The portfolio value of a savings account increased steadily to slightly above 300 thousand at the end of 30 years, which was the principal invested over 30 years.

## Comparisons after Tax

In Table 4, we demonstrate the expected return, standard deviation, and Sharpe ratio after-tax for each investment at different tax rates. As we mentioned above, the expected return and standard deviation of PRIRAs after tax are the same as that before tax. In the second column of Table 4, 0% tax rate corresponds to before tax values of the statistics. As we can see, VFINX had the largest positive after-tax expected return (about 11%), and the savings account had the lowest (0.225%). The annual expected return of the diversified portfolio after tax was about 9%, which was approximately 2% lower than that of VFINX, yet 3 times higher than the average inflation. Moreover, the annual expected return of VBMFX after tax was about 5%. Clearly, the annual expected returns of PRIRAs after tax was lower than that of the VFINX and the diversified portfolio. The expected returns of VBMFX after-tax (taxed at 7% and 14%) were greater than that of $PRIRA_1$. Interestingly, the annual expected return of VBMFX after-tax (taxed at 25% and 33%) underperformed PRIRA$_1$. The expected return of VBMFX after-tax (taxed at 33%) underperformed PRIRA$_2$.

**Table 4 After-tax Comparisons for Different Investment Instruments**

|  | Tax Rate | E(R) | σ(R) | Sharpe ratio |
|---|---|---|---|---|
| **PRIRA$_1$** | / | 4.7% | 8.5% | -14.8% |
| **PRIRA$_2$** | / | 4.45% | 4.9% | -30.82% |
| **Savings Account** | 0% | 0.25% | / | / |
|  | 10% | 0.225% | / | / |
| **VFINX** | 0% | 11.9% | 17% | 34.9% |
|  | 7% | 11.7% | 17% | 33.7% |
|  | 14% | 11.6% | 17% | 33.2% |
|  | 25% | 11.3% | 17% | 31.4% |
|  | 33% | 11.1% | 17% | 30.2% |
| **VBFMX** | 0% | 6% | 4.7% | 0.85% |
|  | 7% | 5.6% | 4.7% | -7.6% |
|  | 14% | 5.2% | 4.6% | -16.5% |
|  | 25% | 4.6% | 4.5% | -30.2% |
|  | 33% | 4.2% | 4.5% | -39.1% |
| **Diversified Portfolio** | 0% | 10% | 12.2% | 33% |
|  | 7% | 9.8% | 12.2% | 31.4% |
|  | 14% | 9.6% | 12.1% | 30% |
|  | 25% | 9.2% | 12.1% | 26.7% |
|  | 33% | 9% | 12.1% | 25% |

The fourth column presents a small tax effect on the risk volatility of VBMFX and the diversified portfolio. The standard deviations of VBMFX and the diversified portfolio decreased when the tax rates increased. On the other hand, the tax had no effect on the standard deviation of VFINX.

In the fifth column, the Sharpe ratio of VFINX, VBMFX, and the diversified portfolio decreased with the increased tax rates. It indicated that with the increased tax rate, VFINX, VBMFX, and the diversified portfolio achieved less return per each unit of risk. Despite the reduction in expected returns in the VFINX and diversified portfolios, the Sharpe ratio of VFINX and the diversified portfolio were positive and far greater than that of $PRIRA_s$. Note that the Sharpe ratios of VBMFX became negative after being subject to the tax. Interestingly, the Sharpe ratios of VBMFX, which were subjected to 14%, 25%, and 33% ordinary income tax, were less than that of $PRIRA_1$. However, a negative Sharpe ratio is difficult to evaluate because a negative excess return with a large standard deviation will make the Sharpe ratio less negative. For example, the negative excess return of $PRIRA_1$ (4.7%-5.96%=-1.26%) was similar to that of VBMFX if VBMFX were subjected to a 14% tax (5.2%-5.96%=-0.76%), but the standard deviation of $PRIRA_1$ was about two times than that of VBMFX. Thus, the Sharpe ratio of $PRIRA_1$ was greater than that of VBMFX (the Sharpe ratio of $PRIRA_1$ was less negative), which indicates that $PRIRA_1$ performed better than VBMFX.

In Table 5, we present the expected portfolio value based on different ordinary income tax rates after 10 years and 30 years, respectively. Keep in mind that the interest income tax rate is 10%.

**Table 5 Expected Portfolio Value of Each Investment after Tax after 10 years and 30 years based on Different Tax Rate**

| | | | Ordinary Income Tax Rate=7% | | | |
|---|---|---|---|---|---|---|
| Time Period | PRIRA$_1$ | PRIRA$_2$ | Savings Account | VFINX | VBMFX | Diversified Portfolio |
| 10 years | 129,861.11 | 128,052.16 | 101,245.89 | 193,198.18 | 136,602.02 | 173,323.50 |
| 30 years | 660,820.95 | 631,847.13 | 310,693.69 | 2,543,670.37 | 778,355.01 | 1,739,129.88 |
| | | | Ordinary Income Tax Rate=14% | | | |
| Time Period | PRIRA$_1$ | PRIRA$_2$ | Savings Account | VFINX | VBMFX | Diversified Portfolio |
| 10 years | 129,861.11 | 128,052.16 | 101,245.89 | 192,095.43 | 133,561.21 | 171,358.81 |
| 30 years | 660,820.95 | 631,847.13 | 310,693.69 | 2,492,793.06 | 723,422.85 | 1,671,728.43 |
| | | | Ordinary Income Tax Rate=25% | | | |
| Time Period | PRIRA$_1$ | PRIRA$_2$ | Savings Account | VFINX | VBMFX | Diversified Portfolio |
| 10 years | 129,861.11 | 128,052.16 | 101,245.89 | 188,826.08 | 129,134.28 | 167,498.79 |
| 30 years | 660,820.95 | 631,847.13 | 310,693.69 | 2,346,463.82 | 649,053.13 | 1,545,153.55 |
| | | | Ordinary Income Tax Rate=33% | | | |
| Time Period | PRIRA$_1$ | PRIRA$_2$ | Savings Account | VFINX | VBMFX | Diversified Portfolio |
| 10 years | 129,861.11 | 128,052.16 | 101,245.89 | 186,678.56 | 126,270.09 | 165,602.93 |
| 30 years | 660,820.95 | 631,847.13 | 310,693.69 | 2,253,944.55 | 604,317.63 | 1,485,752.17 |

Based on different tax rates, the expected portfolio values of VFINX after-tax were highest among six investments after 10 years and 30 years. The expected portfolio values of the diversified portfolio were comparably smaller than that of VFINX at the same tax rates, yet markedly higher than the PRIRA$_s$. The expected portfolio values of VBMFX were slightly higher than that of PRIRA$_1$ at 7% and 14% ordinary income tax rates. However, PRIRA$_1$ outperformed the VBMFX at 25% and 33% tax rates. The expected portfolio values of VBMFX were slightly higher than that of PRIRA$_2$ at 7%, 14% and 25% ordinary income tax rates. However, PRIRA$_2$ outperformed the VBMFX at 33% tax rate.

Figure 6 and Figure 7 present the comparison about the expected portfolio value based on different tax rates over 10 years and 30 years, respectively.

**Figure 3 The Expected Portfolio Value of Each Investment over 10 Years based on Different Tax Rates**

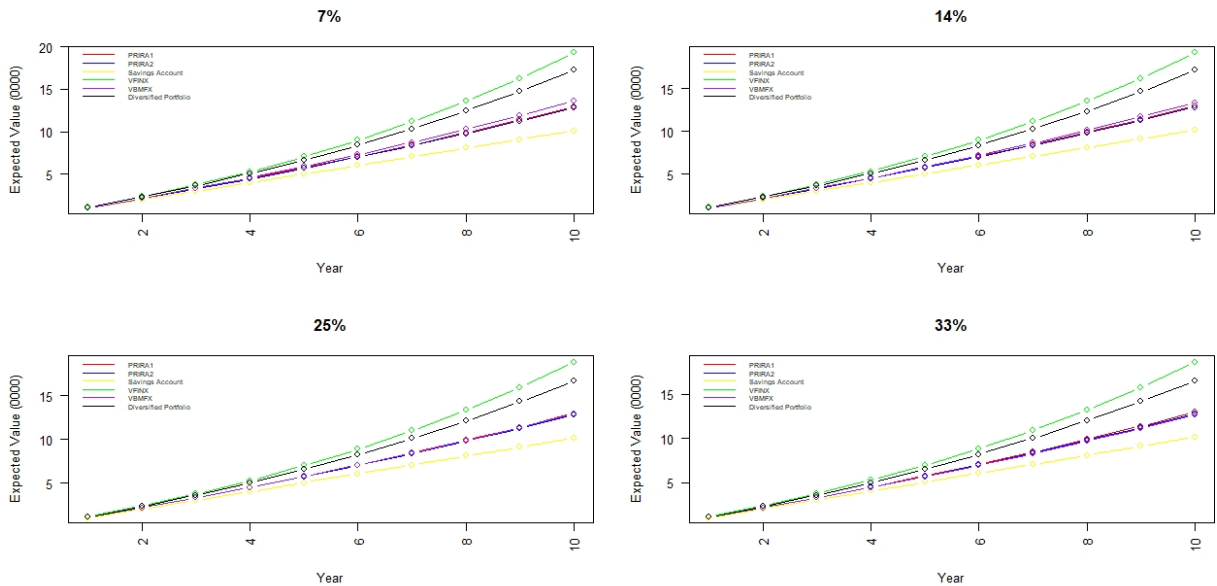

Note that, the expected portfolio value of PRIRA$_s$ (red line and blue line) were slightly below that of the VBMFX (purple line) at 7%, and 14% tax rate. Moreover, the curves of the expected portfolio value of PRIRA$_s$ and VBMFX were almost overlapping at the 25% income tax rate. However, the expected portfolio value of PRIRA$_s$ gradually surpassed that of VBMFX at the 33% tax rate.

**Figure 4 The Expected Portfolio Value of Each Investment over 30 Years based on Different Tax Rates**

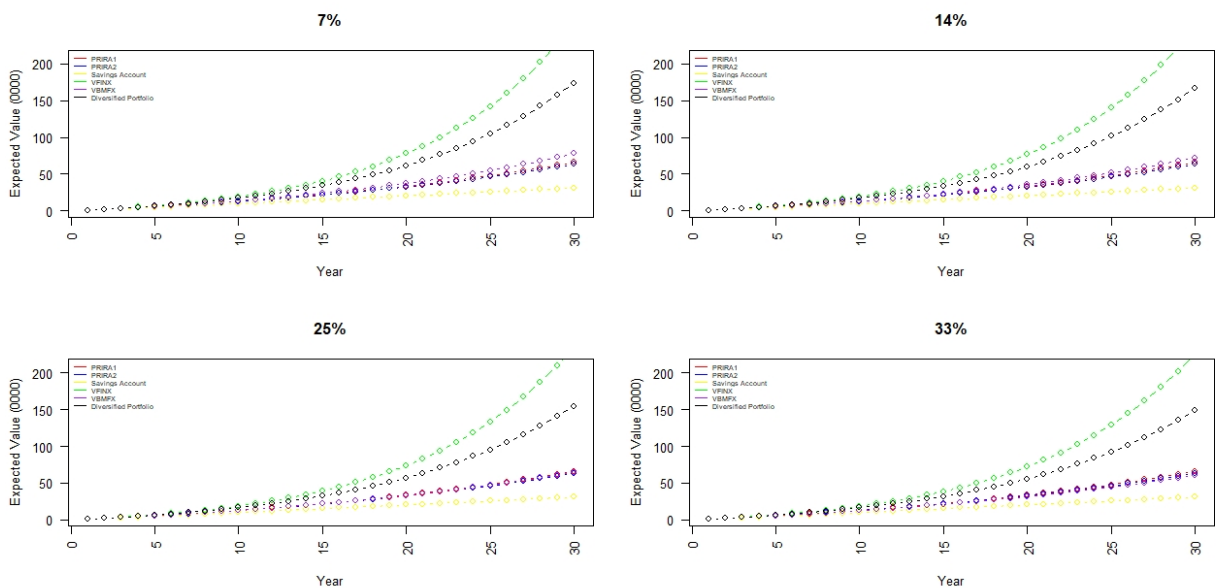

## Discussion

This research aimed to estimate the expected return and risk of the Puerto Rico IRA (PRIRAs) and compare their statistical properties to other common investing strategies. In order to examine this, the law of iterated expectation and the law of total variance were used to construct the volatility model. The comparisons between PRIRAs and the benchmarks also accounted for the tax impact on the expected return and volatility.

The volatility model estimated the annual expected return and standard deviation for $PRIRA_1$ (E(R)=4.7%; σ(R)=8.5%) and $PRIRA_2$ (E(R)=4.45%; σ(R)=4.9%).

This work has shown that investing in the stock market or a diversified portfolio with stocks and bonds has a higher expected return than $PRIRA_s$. While accompanied by higher risks than $PRIRA_s$, the latter alternatives showed expected returns lower than the risk-free rate thout $PRIRA_1$ outperformed investing in bonds after tax at 25% and 33% ordinary income tax rate. Although the expected return of PRIRA$_1$ was slightly greater than that of investing in bonds, which was taxed at the high tax rates (25% and 33%), the volatility of PRIRA$_1$ return was almost two times that of investing in bonds. The expected return of PRIRA$_2$ outperformed investing in bonds after tax at 33% ordinary income tax rate, while PRIRA$_2$ had greater risk. The expected return of PRIRA$_2$ was slightly lower than that of PRIRA$_1$ with less investment risk.

Our findings suggest that, at best, PRIRA$_s$ may be reasonable for some risk-averse investors thanks to the principal protection and tax deferral. After considering tax effects, PRIRA$_1$ may provide a higher expected return than the bonds index funds to some investors, albeit suffering about two times volatility than that of bonds index funds. Generally, PRIRA$_2$ may be a good investment option for investing purposes with high expected return and reasonable investing risk.

There are some limitations to this research. When analyzing the properties of PRIRA$_2$, we arbitrarily used the 3-year credit method, which gave us the best results for the expected return and risk. If we use another credit method as the benchmark, we may change our investing recommendation. In considering the tax effects process, we assumed that the investor didn't withdraw the money from the account. That said, the work clearly illustrates the statistical properties of PRIRAs, assuming the investor invests in the traditional IRA, but it also raises the question of the statistical properties of PRIRAs based on the Roth IRA. Furthermore, we considered the tax effects on different investments, just assuming the investor was taxed at Puerto Rican local taxes. Although most Puerto Ricans do not have to pay the federal income tax, some people in Puerto Rico file (jointly) federal taxes, not state taxes. These people have the option to have federal IRA, therefore not having to pay local taxes. In such cases, a strategy of investing in stocks, bonds, or a diversified portfolio would be much more beneficial than PRIRAs; the latter would make little sense as an investment vehicle.

Future research includes analyzing the statistical properties of federal IRA and compare the differences between federal IRA and the local IRA. Local institutions may consider modifying the current retirement investment products to satisfy local investors' needs. They may also consider creating more diversified investment vehicles for local investors not only for retirement savings but also for investing purposes. Local government should improve residents' awareness of saving money for retirement, which can greatly decrease the future burden of the local retirement system. The Puerto Rican government should reconsider modifying the current investment law to facilitate a diversified economy and incentivize local investment companies to construct more appealing investment products.